\documentclass{aa} 

\usepackage{txfonts}
\usepackage{natbib}
\usepackage{graphicx}
\usepackage{color}
\usepackage{soul}

\bibpunct{(}{)}{;}{a}{}{,}

\title{FIRST, a fibered aperture masking instrument}
\subtitle{I. First on-sky test results}

\author{
   E. Huby\inst{1}
\and G. Perrin \inst{1}
\and F. Marchis \inst{2,3}
\and S. Lacour \inst{1}
\and T. Kotani \inst{4}
\and G. Duch\^ene \inst{3,5}
\and E. Choquet \inst{1}
\and E. L. Gates \inst{6}
\and J. M. Woillez \inst{7}
\and O. Lai \inst{8}
\and P. F\'edou \inst{1}
\and C. Collin \inst{1}
\and F. Chapron \inst{1}
\and V.~Arslanyan \inst{1}
\and K.~J.~Burns \inst{2,3}
}
\institute{
     LESIA, Observatoire de Paris, CNRS, UPMC, Universit\'e Paris-Diderot, Paris Sciences et Lettres, 5 place Jules Janssen, 92195 Meudon, France
\and Carl Sagan Center at the SETI Institute, 189 Bernardo Av., Mountain View CA 94043, USA
\and Department of Astronomy, University of California at Berkeley, Hearst Field Annex, B-20, Berkeley CA 94720-3411, USA
\and ISAS/JAXA, 3-1-1 Yoshinodai, Chuo-ku, Sagamihara 252-5210 Japan
\and UJF-Grenoble 1 / CNRS-INSU, Institut de Plan\'etologie et d'Astrophysique (IPAG) UMR 5274, Grenoble, F-38041, France
\and University of California Observatories/Lick Observatory, P.O. Box 85, Mount Hamilton, CA 95140, USA
\and W. M. Keck Observatory, 65-1120 Mamalahoa Hwy, Kamuela, HI 96743, USA
\and Canada-France-Hawaii Telescope, 65-1238 Mamalahoa Hwy, 96743 Kamuela, HI, USA
}

\abstract{}{In this paper we present the first on-sky results with the fibered aperture masking instrument FIRST. Its principle relies on the combination of spatial filtering and aperture masking using single-mode fibers, a novel technique that is aimed at high dynamic range imaging with high angular resolution.}
{The prototype has been tested with the Shane 3--m telescope at Lick Observatory. The entrance pupil is divided into sub-pupils feeding single-mode fibers. The flux injection into the fibers is optimized by a segmented mirror. The beams are spectrally dispersed and recombined in a non-redundant exit configuration in order to retrieve all contrasts and phases independently.}
{The instrument works at visible wavelengths between 600\,nm and 760\,nm and currently uses nine of the 30 43\,cm sub-apertures constituting the full pupil. First fringes were obtained on \object{Vega} and Deneb. Stable closure phases were measured with standard deviations on the order of 1 degree. Closure phase precision can be further improved by addressing some of the remaining sources of systematic errors. While the number of fibers used in the experiment was too small to reliably estimate visibility amplitudes, we have measured closure amplitudes with a precision of 10\,\% in the best case.}
{These first promising results obtained under real observing conditions validate the concept of the fibered aperture masking instrument and open the way for a new type of ground-based instrument working in the visible. The next steps of the development will be to improve the stability and the sensitivity of the instrument in order to achieve more accurate closure phase and visibility measurements, and to increase the number of sub-pupils to reach full pupil coverage.}


\begin{document}

\maketitle

\section{Introduction}

The search for faint objects such as exoplanets plays an essential role in the development of high angular resolution techniques. Indeed, the detection of companions orbiting around a parent star requires high resolution on one hand, because of the small separation between them, and high dynamic range on the other hand, because of the substantial flux ratio. More precisely, the contrast between the companion and the star can easily reach values from 10$^{-3}$--$\,10^{-5}$ \citep[for hot Jupiters in the visible, see][]{Demory2011} down to 10$^{-8}$ or less depending on the wavelength range and the size and distance of the planet. This kind of target is particularly challenging for ground-based telescopes that are subject to atmospheric effects, the main cause of image degradation. In ideal conditions, without atmospheric turbulence, the angular resolution of a diffraction-limited telescope is $\lambda / D$, with $\lambda$ the working wavelength and $D$ the pupil diameter of the telescope. In presence of atmospheric turbulence, however, the angular resolution is ruled by the Fried parameter $r_0$, the equivalent diameter of a telescope working in the diffraction-limited regime. \\
\indent Adaptive Optics (AO) systems can restore the diffraction limit by actively correcting the phase fluctuations using a deformable mirror \citep{Rousset1990}. Current AO systems are still limited by speckle noise within the central arcsecond \citep{Guyon2005}, and are difficult to implement at optical wavelengths. Additional data reduction is necessary to reach high dynamic ranges at small separations. For instance, the angular differential imaging technique has been applied to detect planets around HR\,8799 \citep{Marois2008, Marois2010}, and saturated image subtraction has been used for the detection of a planet around $\beta$ Pictoris \citep{Lagrange2010}. Both achieve 4\,$\sigma$ planet-to-star flux ratios of 10$^{-4}$ to $5\times 10^{-5}$ (in $\mathrm{K}_\mathrm{s}$ and L' bands) at separations of a few $\lambda/D$ (i.e.\ a few tenths of an arcsecond).\\
\indent In order to reach high dynamic ranges, the AO technique has also been coupled with coronagraphic instruments that aim to block the light coming from the star with a mask. Future instruments in development should provide contrast ratios on the order of 10$^{-5}$, down to 10$^{-7}$ for the brightest targets, at subarcsecond separation, e.g. SPHERE at the VLT \citep{Beuzit2010, Mesa2011}, GPI at Gemini \citep{Macintosh2008}, and HiCIAO at Subaru \citep{Hodapp2008, Suzuki2010}. Impressive results have also been obtained by \citet{Serabyn2010} with one 1.5\,m sub-aperture of the Palomar Hale telescope using a vortex coronograph: they detected the HR\,8799d planet at a separation of $2 \lambda/D$ in $\mathrm{K}_\mathrm{s}$ band and announced a 4\,$\sigma$ detection limit of a few 10$^{-5}$ at $\lambda/D$, very close to the photon noise limit. However, this method is difficult to scale to large telescopes, where the ratio $D/r_0$ is much larger.\\
\indent Post processing of very short exposure images also enables the restoration of the diffraction limit. Speckle interferometry \citep{Labeyrie1970} is well suited to detecting binaries but provides restricted dynamic range at short separations. Aperture masking \citep{Haniff1987} is another solution exploiting short exposure images obtained through a mask with holes (sub-apertures) placed in the pupil plane in order to remove the noise due to atmospheric perturbations. This technique has established itself as the standard for high resolution imaging with contrast ratios on the order of 10$^{-3}$ at $\lambda/D$ \citep{Hinkley2011}. The drawback of this method is that only a small fraction of the telescope pupil can be used, making photon noise an important factor in dynamic range limitation. This can only be overcome for bright sources where dynamic range is primarily limited by calibration accuracy. Once calibration issues are solved, higher dynamic ranges can be contemplated by increasing the number of collected photons. An alternative has been tested at Keck Observatory by dividing the pupil into sets of non-redundant configurations \citep{Monnier2009}. Differential tilts of the segmented pupil allow the images of the sets to be focused at different locations in the image plane. This enables the use of the whole pupil along with the aperture masking technique.\\
\indent Among the techniques developed to mitigate atmospheric phase fluctuations, the closure phase technique \citep{Jennison1958} has to be mentioned and highlighted. It consists of combining the phases of three different baselines that form a closed triangle, and is, by definition, unaffected by phase shifts induced by atmospheric turbulence. The closure phase is a widely used observable quantity in aperture masking \citep{Lacour2011} and long baseline interferometry \citep{Zhao2011}. This technique seems even more promising since \citet{Martinache2010} has demonstrated that in the case of a high Strehl ratio, closure phase-like quantities called Ker-phases can be measured from direct images. In this work, they report on the detection of a known companion from Hubble Space Telescope data with a $5\times 10^{-3}$ contrast ratio at $\lambda / D$ and derive a detection limit of $2\times10^{-2}$ at $0.5\,\lambda/D$ with a 99\,\% confidence level.\\
\indent In this context, \citet{Perrin2006} proposed the concept of a fibered aperture masking instrument called FIRST for \emph{Fibered Imager for Single Telescope} to achieve high dynamic ranges down to a fraction of the diffraction limit. This imager combines the techniques of aperture masking and spatial filtering using single-mode fibers. The fibers only transmit a coherent area of the incident wavefront, resulting in a plane wavefront at the fiber end. The spatial phase fluctuations are thus converted into intensity variations and only residual differential piston terms between the sub-pupils remain. Moreover, optical fibers fulfill another function: the entrance pupil of the telescope is divided into sub-pupils and rearranged into a non-redundant configuration via the fibers. This non-redundancy is widely used in aperture masking because all fringe contrasts and phases can be retrieved independently without the attenuation observed when complex vectors with random phases are added, as in the case of a full pupil. In principle, the combination of single-mode fibers and non-redundancy allows the perfect calibration of the point spread function over the reconstructed field of view (to within photon and detector noise in an exposure time smaller than the coherence time of turbulence) and the elimination of speckle noise to reach dynamic ranges of 10$^{-6}$ at a fraction of $\lambda /D$ \citep{Lacour2007}. The FIRST concept is therefore complementary to AO, which allows even higher dynamic ranges but at longer distances from the central object.\\
\indent A prototype of this instrument was developed and tested at Paris Observatory \citep{Kotani2009}. Laboratory experiments have shown that the image of a simulated binary object could be retrieved from the closure phases and visibilities measurements. For the sake of simplicity, this prototype uses 9 out of 36 sub-pupils, or one quarter of the total pupil area, and works in the visible (600\,nm\,--\,800\,nm). Working in the visible provides good resolution (50\,mas at 700\,nm with a 3--m\,telescope at the diffraction limit) and, although the infrared wavelengths are more favorable to the detection of low mass companions, could eventually offer complementary information by covering a larger spectral range.\\
\indent In this paper, we present the first on-sky results obtained in July 2010 with the same prototype set up on the Shane 3--m telescope at Lick Observatory. In the next section the instrument and the key technologies are described. The measurement procedure during the observing nights is detailed in Sect.~3 and the data reduction method is explained in Sect.~4. Section~5 presents the first closure phase and closure amplitude measurements obtained on \object{Vega} and Deneb. These results and future improvements are discussed in Sect.~6. Conclusions are drawn in Sect.~7.

\section{Description of the instrument}

\subsection{Telescope interface} 
The instrument was mounted at the Cassegrain focus of the Shane 3--m telescope at Lick Observatory. A special interface was constructed to mount the 1.3\,m $\times$ 0.6\,m optical bench, consisting of five 50\,cm long posts screwed on the bench of the AO system and supporting the whole instrument. A beam splitter cube, positioned in the light path of the AO system just before the wavefront sensor, sends 50\,\% of the flux from the telescope to the FIRST instrument, passing through a 10\,cm diameter hole in the bench.\\

\subsection{Optical set-up}
\begin{figure}
\centering
\includegraphics[width=8cm]{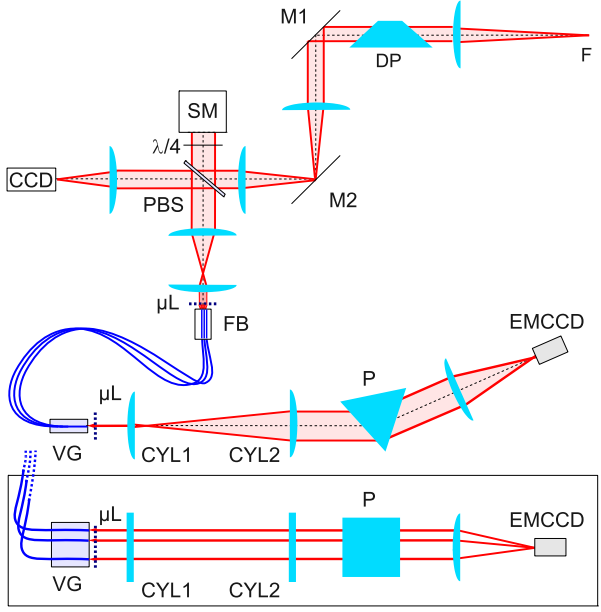}
\caption{Schematic view of the optical set-up. F=telescope focal point. DP=Dove prism. M1=pupil mirror. M2=field mirror. PBS=polarizing beam splitter. $\lambda/4$=quarter wave plate. SM=segmented mirror. $\,\mu$L=microlens array. FB=fiber bundle. VG=v-groove. CYL1, CYL2=cylindrical lenses. P=dispersive prism. \emph{Inset:} side view of the recombination optics. Only three fibers are drawn.}
\label{Setup}
\end{figure}

\begin{figure}
\centering
\resizebox{\hsize}{!}{\includegraphics{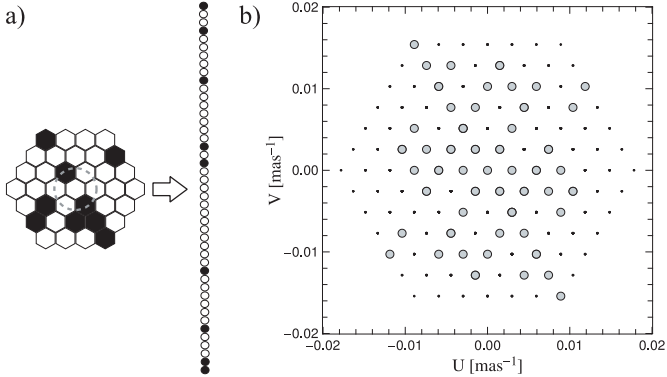}}
\caption{a) Entrance pupil with the nine sub-aperture configuration and the non redundant configuration for recombination. The dashed gray circle indicates the central obstruction of the Shane telescope. Positions of the fibers in line are (from bottom to top): 1, 2, 6, 13, 26, 28, 36, 42 and 45. b) Frequency coverage corresponding to the chosen entrance pupil configuration.}
\label{Pupils}
\end{figure}

Figure~\ref{Setup} is a schematic representation of the optical set-up. The practical implementation is based on a few key components that come from advanced technologies: \\
\indent - the Segmented Mirror consists of 37 700$\,\mu $m-diameter (vertex-to-vertex) hexagonal segments that can be independently positioned in piston/tip/tilt (each segment is controlled by three actuators) in order to adjust the phase and steer beams into individual fibers. This mirror was manufactured by \emph{Iris AO} \citep{Helmbrecht2006, Helmbrecht2011}.\\
\indent - the Fiber Bundle consists of 36 single-mode Polarization Maintaining (PM) fibers (\emph{Nufern PM630-HP}, 570\,nm cut-off wavelength) aligned on a 250$\,\mu $m-pitch hexagonal grid for sampling of the pupil, without a central fiber because of the telescope central obscuration. The first prototype uses 9 out of the 36 available fibers. This bundle was manufactured by \emph{Fiberguide Industries}.\\
\indent - the Microlens Arrays consist of 250$\,\mu$m-diameter fused silica microlenses aligned on the same hexagonal grid to focus the light into the fibers. They were manufactured by \emph{SUSS MicroOptics}.\\
\indent - the v-groove consists of a silicon chip where the PM single-mode fibers are precisely positioned along a line with a 250 $\,\mu$m-pitch, enabling the arrangement of the fibers according to the desired non-redundant exit pupil configuration. It was manufactured by \emph{OZ Optics}.\\
\indent The telescope pupil is divided into sub-pupils by the microlens array. Their numerical aperture and hence focal length has been chosen so that the injection efficiency into the fibers is optimized. The beams corresponding to each sub-aperture are focused on the cores of the single-mode fibers gathered in the fiber bundle. The segmented mirror is placed in a pupil plane and each segment is aligned in order to maximize the injection rate into each corresponding fiber. An afocal combination of two lenses reduces the beam diameters to match the pitch of the fiber bundle (700$\,\mu$m to 250$\,\mu$m).\\
\indent The 9-fiber entrance configuration can be seen on Fig. \ref{Pupils}. The size of the sub-pupils have been chosen in order to maximize the sensitivity (which increases with the size of sub-pupils and hence decreases with their number if the total pupil size is fixed) and optimize the coupling efficiency of the fibers. According to \citet{Lacour2007a}, a diameter of $\sim3r_\mathrm{0}$ is a good compromise. With $r_\mathrm{0}$ on the order of 15\,cm at 700\,nm under good seeing conditions, this leads to sub-pupils of about 45\,cm.\\
\indent The segmented mirror is used at normal incidence to minimize flux losses when the segments are tilted. This is made possible by a beam splitter which is a polarizing cube used together with a quarter wave plate to send half of the flux (one polarization state) towards the fibers. The other half is used to control the position of the object in the field of view (a CCD camera is placed in an image plane). In any case, the two polarization states can not be kept because of the birefringence of the fibers. Indeed, the polarization states will be phase shifted at the PM fiber end, inducing contrast losses. To prevent this, a single polarization state (corresponding to the slow axis of the PM fibers) is selected by the beam splitter cube. With this set-up, 100\,\% of the light of one polarization is therefore used to produce the fringes.\\
\indent The fiber outputs are then rearranged according to a one-dimensional, non-redundant configuration that can be seen in Fig. \ref{Pupils} and is the most compact 9-fiber linear configuration. Each pair of sub-pupils forms a unique baseline. A second microlens array collimates the 9 output beams. The direction perpendicular to the fiber line is dedicated to spectral dispersion. An anamorphic system consisting of the afocal combination of two cylindrical lenses enlarges the beam in the horizontal direction by a factor of 20, enabling higher spectral resolution. The beams are then spectrally dispersed in the orthogonal direction thanks to an equilateral BK7-prism such that the spectral resolution is around 150. The final focusing lens forms the image on the EMCCD (Electron Multiplying Charge Coupled Device) camera (Luca-S EMCCD camera from Andor Technology). The detector consists of 496$\times$658 10\,$\mu$m pixels and the quantum efficiency of the camera is 50\,\% at 600\,nm, 35\,\% at 700\,nm and 20\,\% at 800\,nm. When using the EM mode, the maximum gain is $\times$\,200.\\
\indent The fibers are therefore the core of the instrument, and also the most critical part. Indeed, the total fiber length of each sub-aperture must be equal in order to minimize the optical path differences (OPD) and thus maximize the fringe contrasts. The coherence length gives the maximum OPD where fringes can still be observed, and is 100$\,\mu$m in air and 67$\,\mu$m in glass for a spectral resolution of 150 at $\lambda=700\,$nm. Moreover, equal fiber lengths are necessary to minimize chromatic dispersion due to the frequency-dependent refractive index of the fiber and waveguide effects \citep{Dyer1997}.\\
\indent Note that the Dove prism placed in the collimated beam is not absolutely necessary. It provides a way to cover a larger area of the (u,v) plane with 9 fibers. Indeed, when the prism rotates about the optical axis, the beam rotates at twice the rate of the prism. The (u,v) plane coverage is therefore extended by rotating the initial frequency positions. However, this component will not be needed when the telescope pupil is fully sampled.

\section{Principle of measurements}

The critical parts of the alignment are the injection into the single-mode fibers and re-collimation at the v-groove fiber ends. The core diameter of the fibers is 4$\,\mu$m, while the focal length of the microlens is about 1\,mm. Stable alignments and a systematic procedure during night observations were thus needed.\\

\subsection{Injection optimization procedure} 
\label{sec:opti}
The segmented mirror is the key component for fiber coupling optimization. Software has been developed to control each segment. A raster scan has been implemented to automatically find the best orientation of each segment by measuring the flux transmitted by each fiber onto the EMCCD camera as a function of segment tip-tilt values. This alignment is aimed at correcting the static aberrations introduced by the optical train, while the IR-optimized AO system actively corrects the turbulent fluctuations of the wavefront. At the scale of each sub-pupil, the correction amounts to removing part of tip-tilt while higher order aberrations are left unchanged. In addition, the average differential phase from one pupil to another remains negligible as the AO system reconstructs a continuous wavefront, so the AO system is equivalent to a multi-pupil tip-tilt and differential piston corrector. In addition, a global image alignment was performed with a motorized steering mirror (M1 on Fig.~\ref{Setup}). Pointing errors are monitored on the CCD placed in an image plane. The alignment is adjusted by maintaining the target at a fixed position in the field of view.\\
\indent Details of the instrument throughput are presented in Table \ref{Transmissions}. The total efficiency is deduced from data taken on \object{Vega} by comparing the number of photons expected from the stellar spectrum in the 600\,--\,760\,nm range to the number of counts measured on the detector in 200\,ms. Telescope, AO transmission and FIRST optics transmission have been estimated with the theoretical transmission and reflection coefficients. An upper value of the fiber injection efficiency has been obtained by assuming that injection losses are due to residual tip-tilt, mode mismatch \citep[a 22\,\% loss according to][]{Shaklan1988}, and high order phase perturbations.\\
\indent Tip-tilt data were regularly measured by the AO wavefront sensor during the observations to evaluate the typical performance of the tip-tilt correction. The wavefront sensor consists of sub-pupils of the same diameter as the FIRST sub-pupils (that is, about 43\,cm on the primary mirror). The standard deviation of the residual fluctuations is on the order of 125\,mas per sub-pupil on the sky, or a 1.0\,$\mu$m lateral offset of the Airy pattern on the fiber core. The average injection efficiency due to tip-tilt fluctuations is of 76\,\% at 700\,nm, computed with the formula:
\begin{equation}
\eta=\mathrm{exp}\left(\frac{-2\theta^2 f^2}{\omega_0^2}\right),
\end{equation} 
with $\theta$ the tip-tilt angle on the microlens, $f$ the microlens focal length, and $\omega_0$ the waist of the fiber mode. This expression can be used under the assumption of perfect mode matching between the beam and the fiber mode.\\
\indent At the time of the observations, the typical r$_0$ at 550\,nm was 22\,cm, and hence the spatial phase variance over a sub-pupil was 1.9\,rad$^2$ at 700\,nm. The phase variance (free of tip-tilt) is 0.25\,rad$^2$ which translates to a coherent energy of 78\,\%. Taking into account the mode mismatch and residual tip-tilt, the fiber injection efficiency is 46\,\%.\\
\indent The throughput budget yields a theoretical efficiency of 1.2\,\%, which overestimates the measured efficiency of 0.21\,\% by a factor of $\sim$5.7. This difference may be explained by imperfect alignments and also non-ideal performance of the optics (ageing, dust, etc \dots). For instance, assuming that all 46 surfaces (that are exposed to air) share the same contribution to the total loss and applying an additional efficiency of 97\,\% per optical surface makes the theoretical transmission drop to 0.25\,\%. This may explain the poor efficiency of FIRST during the first tests reported in this paper. This clearly leaves some margin for improvement. It is obvious that the current sensitivity limit can potentially be improved by reducing the number of optical surfaces thanks to a different set-up and by improving the quality of optics. Achieving the theoretical 1\,\% overall efficiency therefore seems a sensible goal and would lead to a sensitivity limit improved by 1 magnitude in the photon noise-limited regime. This is further discussed in Sect.~\ref{sec:discussion}. \\
\indent Note that the data reported here were obtained with only four fibers at sufficiently high signal-to-noise ratios to produce stable fringe patterns (thus leading to stable and relatively accurate closure phase measurements, see Sect.~4). Two out of the nine fibers fell inside the central obstruction of the telescope, as shown in Fig \ref{Pupils} (this will be changed in the future with a new fiber configuration). In addition, one fiber was damaged during transport and did not transmit light. Two other fibers transmitted only low fluxes or produced low contrast fringes (due for example to a high flux ratio between some pairs of sub-apertures, or perhaps to unequal fiber lengths causing the mean OPD to be comparable to the coherence length).
\begin{table}
\caption{Instrument transmission efficiency.}
\label{Transmissions}
\centering
\begin{tabular}{c c}
 Item & Transmission (\,\%) \\
\hline
Telescope and Adaptive Optics & 48\\
FIRST optics				  & 22\\
Entrance beam splitter		  & 50\\
Polarization selection	      & 50\\
Injection rate into fibers	  & 46\\
\hline
Theoretical efficiency		  & 1.2\\
Measured efficiency		      & 0.2\\
\hline
\end{tabular}
\end{table}

\subsection{Image acquisition}
\indent After alignment of the nine segments, data cubes of 50 images were acquired, with integration times of 200\,ms. Although optimized for the infrared, the AO system provides sufficient wavefront corrections to stabilize the phase of the fringes, thus fulfilling the role of a fringe tracker and making longer integration times possible (coherence time in the visible is only on the order of 10\,ms). The choice of integration time is the result of a compromise between sensitivity and fringe contrast loss that has been roughly evaluated during observations. The study of the residual piston between sub-pupils after the correction applied by the AO system is not trivial and needs to be detailed in a longer study that is beyond the scope of this paper.\\
\indent The camera was used in the Electron Multiplying (EM) mode with the highest gain ($\times200$), making the readout noise negligible with respect to the other noise sources (the dark current and the photon noise are amplified, while the readout noise is not). However, the noise is amplified by a factor of $F=\sqrt{2}$ when using the EMCCD mode. This is equivalent to a reduction of the signal-to-noise ratio by a factor of $\sqrt{2}$ \citep{Denvir2003}, while the average value of the signal itself is not affected. Since the dark current is insignificant in the visible, the acquisitions are photon noise-limited, apart from the multiplicative noise factor.\\
\indent A full acquisition was completed in six steps: four sets of images on the object were taken for the different positions of the Dove prism (0\degr, 45\degr, 90\degr~and 135\degr), followed by one set of sky background with telescope offset 30\,arcsecond and a calibration sequence. During this sequence, the flux of each sub-pupil was independently recorded, with all segments but one being deliberately misaligned. This step is aimed at measuring individual fiber images, which are necessary for modeling the interferogram (see Sect.~4).\\
\indent For the first on-sky tests, the target list was limited to bright objects with $R_\mathrm{mag}\leq3$ near zenith as the instrument suffered from mechanical flexure. In this work, we present the results of observations of \object{Vega} (A0V star, V$_\mathrm{mag}$=0.03, R$_\mathrm{mag}$=0.1) and Deneb (A2I star, V$_\mathrm{mag}$=1.25, R$_\mathrm{mag}$=1.14) as a calibrator. The observations were conducted on July 30 2010 from 06:01 UT to 09:45 UT under good seeing conditions, with a visible $r_0$ estimated to be 22\,cm at 550\,nm on average by the Lick-AO system. The observation log is reported in Table \ref{Obs}, indicating the number of data cubes (each containing 50 frames) and the position of the Dove prism.\\

\begin{table}
\caption{Observation log.}
\label{Obs}
\centering
\begin{tabular}{c c c c c}
 Target & Start time (UT)& Acquisition & Dove position (\degr)\\
\hline
Vega & 06:01:17 & 10 datacubes & 45\\
Vega & 06:09:16 & 10 datacubes & 45\\
Vega & 06:25:12 & 10 datacubes & 90\\
Vega & 06:32:06 & 10 datacubes & 90\\
Vega & 06:43:20 & 10 datacubes & 90\\
Vega & 06:51:23 & 10 datacubes & 90\\
Vega & 07:09:11 & 10 datacubes & 0\\
Vega & 07:17:00 & 10 datacubes & 0\\
Vega & 07:33:48 & 10 datacubes & 135\\
Vega & 07:41:31 & 10 datacubes & 135\\
Vega & 07:51:21 & calibration files &\\
Vega & 07:58:21 & sky background &\\
\hline
Deneb & 08:22:20 & 10 datacubes & 135\\
Deneb & 08:31:20 & 10 datacubes & 135\\
Deneb & 08:51:16 & 10 datacubes & 90\\
Deneb & 08:58:35 & 10 datacubes & 90\\
Deneb & 09:06:51 & 10 datacubes & 90\\
Deneb & 09:28:05 & 10 datacubes & 45\\
Deneb & 09:37:44 & 10 datacubes & 0\\
Deneb & 09:44:58 & sky background \\
\hline
\end{tabular}
\end{table}

\section{Data reduction}

\begin{figure}
\centering
\resizebox{\hsize}{!}{\includegraphics{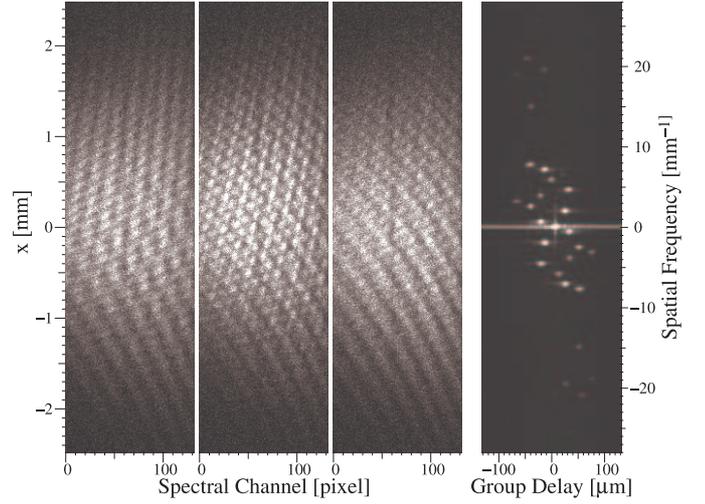}}
\caption{The three images to the left are samples of the acquired data on \object{Vega}, where fringes are particularly visible. The right panel is the 2D Spectral Power Density averaged over 40 data cubes of 50 images each, in a logarithmic scale. One can distinguish 13 peaks, meaning that at least 6 fibers contributed to the interferograms in this particular set of data.}
\label{IMG_Vega}
\end{figure}

Typical images acquired with FIRST are presented in Fig. \ref{IMG_Vega}. The wavelengths are in abscissa and the OPD is in ordinate (presented as the position $x$ on the detector). Each image consists of the 133 spectral channels used for the data reduction. They correspond to the 399 pixels with the highest signal-to-noise ratios in the raw images, binned in sets of 3 to increase the signal-to-noise ratio. The 2D spectral power density shows that the images are the superposition of at least 13 fringe patterns. However, this is not part of the data reduction that is presented in this work, since the data are processed directly in the image plane rather than in the Fourier plane, as for the AMBER long-baseline interferometry instrument \citep{Millour2004}.\\
\indent Indeed, because of the multi-axial beam combination scheme with spatial coding and spectral dispersion, the final interferogram is a linear combination of fixed patterns with parameters linked to injected fluxes, fringe contrasts and phases. If the carrying waves are well determined, the fringe parameters can then be measured through model fitting. Using the formalism of \citet{Lacour2007}, the observables are the complex contrasts $\mu_{ij}$ for each pair of sub-pupils $i$ and $j$:
\begin{equation}
\label{eq:muij}
\mu_{ij}=V_{ij}G_iG_j^*=\vert V_{ij} \vert \vert G_i \vert \vert G_j \vert \mathrm{e}^{i(\Phi_j - \Phi_i + \varphi_{ij})},
\end{equation}
with $V_{ij}$ the complex visibility (modulus $\vert V_{ij} \vert$ and phase $\varphi_{ij}$ ) and $G_i$ the complex transmission in the pupil $i$ (modulus $\vert G_i \vert$, the transmitted flux, and phase $\Phi_i$). The $\mu_{ij}$ are basically the measured Fourier components, while the $V_{ij}$ are the intrinsic Fourier components of the object. The phases of the complex transmissions $G_i$ account for residual differential piston (including atmospheric and instrumental OPD).

\subsection{Model for the interferograms}
The acquired data are spectrally dispersed. Each column of the image corresponds to one spectral channel of width $\delta\lambda$. The intensity recorded for the channel at wavenumber $\sigma$ is:

\begin{eqnarray} \nonumber
\mathcal{I}(x)&=& \sum\limits_{i=1}^{N_{\mathrm{pup}}} \vert G_i\vert^2 a_i(x) \\ 
              &+& 2 \sum\limits_{j<i<N_{\mathrm{pup}}}^{n_\mathrm{B}} \sqrt{a_i(x)a_j(x)}~\Re\left( \mu_{ij}~\mathrm{e}^{ 2 i \pi \sigma B_{ij} \frac{x}{D} } \right) ,
\end{eqnarray}
where $x$ is the position on the detector ($x$ takes discrete values $x_k$, multiples of the pixel size of 10$\,\mu$m), $N_{\mathrm{pup}}$ is the total number of sub-pupils, $n_\mathrm{B}$ the number of baselines, and $D$ the distance to the image plane. $a_i(x)$ is the normalized envelope function for the $i^{\mathrm{th}}$ pupil such that $\sum\limits_{k=1}^{n_\mathrm{P}} a_i(x_k)=\vert G_i\vert$, $n_\mathrm{P}$ being the number of pixels, and $B_{ij}$ is the baseline formed by the pupils $i$ and $j$ in the exit pupil.\\
\indent This expression can also be written:
\begin{eqnarray}\nonumber
\mathcal{I}(x) &=& \sum\limits_{i=1}^{N_{\mathrm{pup}}} \vert G_i\vert^2 a_i(x)\\ \nonumber
               &+& 2 \sum\limits_{j<i<N_{\mathrm{pup}}}^{n_\mathrm{B}} \Re\left(\mu_{ij}\right) \sqrt{a_i(x)a_j(x)} \cos \left(2\pi f_{ij}x\right) \\
               &-& 2 \sum\limits_{j<i<N_{\mathrm{pup}}}^{n_\mathrm{B}} \Im\left(\mu_{ij}\right) \sqrt{a_i(x)a_j(x)} \sin \left(2\pi f_{ij}x\right),
\end{eqnarray}
\noindent with $f_{ij}=\sigma \frac{B_{ij}}{D}$, the spatial frequency corresponding to the baseline $ij$.
\indent Assuming the photometric fluxes $\vert G_i \vert $ and the envelope functions $a_i(x)$ are known, the interferogram can be corrected from the fixed component and becomes:
\begin{eqnarray} \nonumber
\mathcal{I'}(x)&=& \mathcal{I}(x) - \sum\limits_{i=1}^{N_{\mathrm{pup}}}\vert G_i \vert^2 a^i(x) \\
               &=& \sum\limits_{j<i<N_{\mathrm{pup}}}^{n_\mathrm{B}} \Re\left(\mu_{ij}\right) c_{ij}(x) - \sum\limits_{j<i<N_{\mathrm{pup}}}^{n_\mathrm{B}} \Im\left(\mu_{ij}\right) s_{ij}(x),
\end{eqnarray}
with
\begin{eqnarray}
c_{ij}(x)&=& 2\sqrt{a_i(x)a_j(x)} \cos \left(2 \pi f_{ij} x \right),\\
s_{ij}(x)&=& 2\sqrt{a_i(x)a_j(x)} \sin \left(2 \pi f_{ij} x \right).
\end{eqnarray}
$c_{ij}$ and $s_{ij}$ are defined to be the \emph{carrying waves} that constitute the model of the fringe pattern. They have to be precisely calibrated in order to retrieve the real and imaginary parts of the $\mu_{ij}$ by fitting the fringes with this model.
\subsection{Calibration}
The envelope functions of each fiber, $a_i(x)$, are defined from the calibration data described in the previous section. The spatial frequencies, that depend on the wavenumber $\sigma$ and the baseline $B_{ij}$, also have to be determined. Assuming that the baselines are known (the v-groove pitch is given with a 0.5$\,\mu$m precision), a first approximation of the central wavelength of each channel is made using one spectral absorption line observed in the stellar spectrum ($H_\alpha$ at 656\,nm) and two lines due to the atmosphere (O$_2$ at 687\,nm and 760\,nm). More accurate values of the wavenumbers and baselines are obtained by fitting interferograms in each spectral channel. The resulting wavelengths are finally fitted with a $5^\mathrm{th}$ order polynomial function over the spectral channels.

\subsection{Fringe fitting} 
\indent Once this calibration is done, the problem is linear regarding the frequency components and can be expressed in terms of matrices. For one spectral channel, the interferogram is sampled by the number of pixels $n_\mathrm{P}$:

\begin{equation}
\left(
\begin{array}{c}
\mathcal{I}(x_1) \\ \vdots  \\ \mathcal{I}(x_{n_\mathrm{P}}) 
\end{array}
\right) =  M \left(
\begin{array}{c}
\Re(\mu_{ij}) \\ \vdots \\ \Im(\mu_{ij}) \\ \vdots 
\end{array}
\right)
~~\Rightarrow~~
\left(
\begin{array}{c}
R_{ij} \\ \vdots \\ I_{ij} \\ \vdots 
\end{array}
\right) = M^{-1} \left(
\begin{array}{c}
\mathcal{I}(x_1) \\ \vdots  \\ \mathcal{I}(x_{n_\mathrm{P}}) 
\end{array}
\right),
\end{equation}

\noindent with $M$ a $n_\mathrm{P} \times2\,n_\mathrm{B}$ matrix defined by the carrying waves:
\begin{equation}
M = \left( 
\begin{array}{cccccc}
c^1_1  & \cdots & c^{n_\mathrm{B}}_1 & s^1_1 & \cdots & s^{n_\mathrm{B}}_1  \\
\vdots &        &  \vdots   & \vdots &       & \vdots \\
c^1_{n_\mathrm{P}} & \cdots & c^{n_\mathrm{B}}_{n_\mathrm{P}} & s^1_{n_\mathrm{P}} & \cdots & s^{n_\mathrm{B}}_{n_\mathrm{0}}
\end{array}
\right).
\end{equation}

\indent $R_{ij}$ and $I_{ij}$ are the best fit parameter sets in the least-squares sense.\\
\indent This simplified matrix inversion can be written under the assumptions of uncorrelated pixels and constant variance over all pixels. In practice, the matrix $M$ is rectangular and the generalized inverse matrix is determined via singular value decomposition. According to the formalism introduced by \citet{Millour2004}, the matrix $M$ containing the carrying waves is called the \emph{visibility to pixel matrix} (V2PM), while the inverse of this matrix is called the \emph{pixel to visibility matrix} (P2VM).\\

\subsection{Measured quantities}
\label{sec:Meas}
\indent Fringe contrasts $\mu_{ij}$ are measured with: 
\begin{equation}
\label{eq:Muij_fin}
\vert \mu_{ij} \vert =\sqrt{R_{ij}^2+I_{ij}^2}.
\end{equation}
\indent According to Eq. \ref{eq:muij}, the individual fluxes $\vert G_{i} \vert$ are needed to retrieve the object visibility, often measured through the squared visibility:
\begin{equation}
\label{eq:Vij}
\vert V_{ij} \vert^2=\frac{\vert \mu_{ij} \vert^2}{\vert G_{i} \vert^2 \vert G_{j} \vert^2}.
\end{equation}

\indent Closure phase is by definition the phase of the bispectrum $\mu_{ijk}$:
\begin{equation}
\label{eq:Muijk}
\mu_{ijk}=\langle \mu_{ij} \mu_{jk} {\mu_{ik}}^* \rangle,
\end{equation}
\noindent with $ijk$ referring to the pupils forming the considered baseline triangle and $\langle \rangle$ indicating the average over all measurements. The most interesting property of this quantity is that it is independent of differential pistons, which naturally cancel each other out when summing the phases of three baselines that form a closed triangle:
\begin{equation}
\label{eq:CPijk}
CP_{ijk}=arg(\mu_{ijk})=\varphi_{ij}+\varphi_{jk}-\varphi_{ik},
\end{equation}
with $\varphi_{ij}$ the phase of the complex visibility $V_{ij}$.\\
\indent Since the squared visibility and closure phase are expressed as a power law of a noisy quantity (additive photon noise), undesired photon biases affect these estimates. High precision measurements require bias-free estimators (see \citet{Goodman1976} for squared visibility and \citet{Wirnitzer1985} for bispectrum.) Such estimators have not been used for the reduction of the first data. Indeed, only closure phases of unresolved targets have been retrieved for now and in this case photon biases cannot be the cause of non-zero closure phases (the bias is necessarily a positive real number). Concerning the visibilities, Eq. \ref{eq:Vij} states that the individual fluxes of each sub-pupil are needed to estimate the visibility terms. However, the FIRST instrument does not include any photometric channel to measure these quantities. \citet{Lacour2007} developed an algorithm to retrieve object visibilities directly from the $\mu_{ij}$ measurements. This is possible if the rank of the matrix linking the observables ($\mu_{ij}$) and the unknowns (G$_i$ and $V_{ij}$) is at least equal to the number of unknowns, that is, if the entrance pupil configuration shows a sufficient number of redundancies (which is the case for the chosen pupil configuration illustrated in Fig. \ref{Pupils}).\\
\indent For the first-light observations, only four fibers could be used to measure closure phases with reasonable precisions (see Sect.~3.1), reducing the pupil to a non-redundant configuration and making the problem ill-posed (12 complex equations for 6 complex visibility terms and 4 complex transmission terms). Calibrated fringe contrasts could therefore not be measured from the first run data.\\
\indent Another quantity can be built from the uncalibrated contrasts in the same spirit as closure phase:
\begin{equation}
CA_{ijkl}=\frac{\vert\mu_{ij}\vert\vert\mu_{kl}\vert}{\vert\mu_{ik}\vert\vert\mu_{jl}\vert}=\left| \frac{ V_{ij} V_{kl}}{ V_{ik} V_{jl}}\right|.
\end{equation}
This quantity is called closure amplitude (CA) and only depends on the moduli of the visibilities \citep{Readhead1980}.

\section{First results} 

\subsection{Closure Phase results}

\begin{figure*}
\centering
\includegraphics[width=17cm]{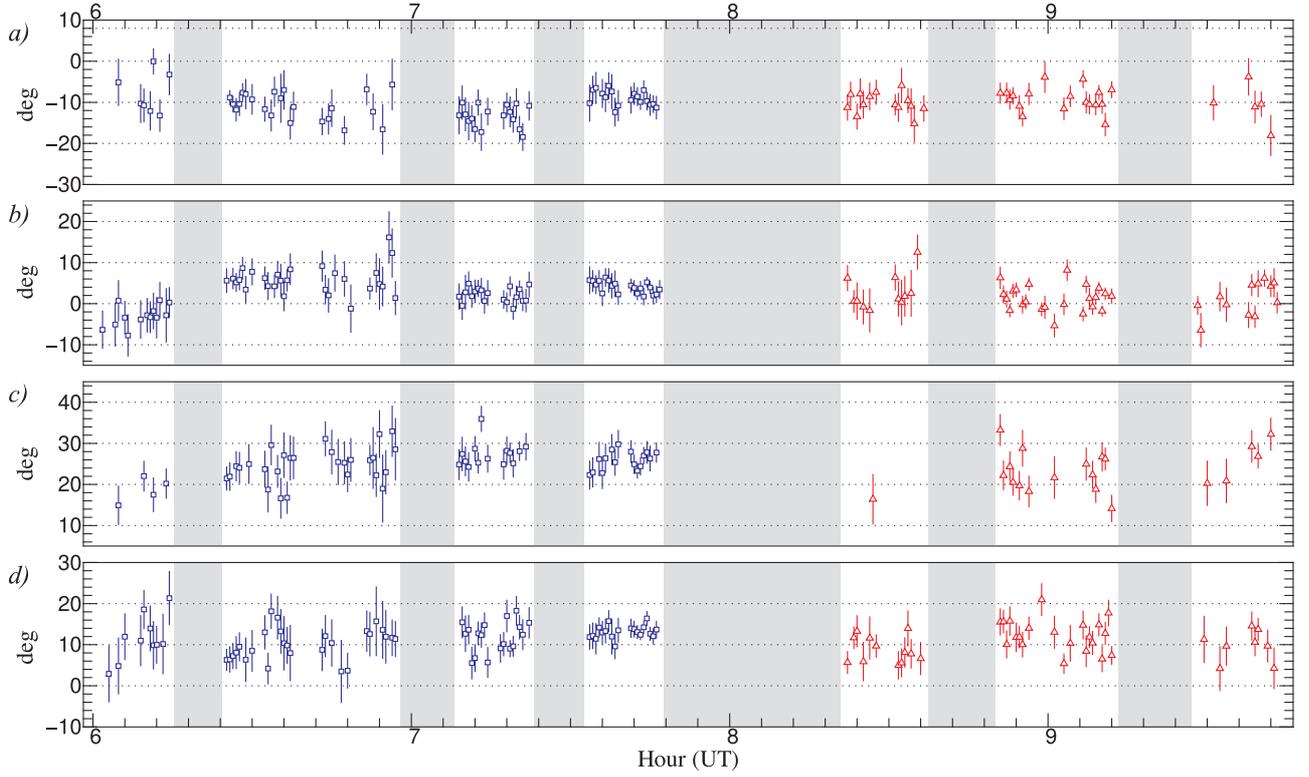}
\caption{Estimates of the closure phase of \object{Vega} (blue squares) and Deneb (red triangles) with rescaled error bars, corresponding to the contributions of four sub-pupils numbered 1, 2, 6 and 13: \emph{a)}: 1-2-6 ; \emph{b)}: 1-2-13 ; \emph{c)}: 1-6-13 ; \emph{d)}: 2-6-13. Each estimate corresponds to 10\,s of observation (average of 50 images of 200\,ms integrations). The error bars represent the standard deviation of the mean over the 50 frames (i.e. the frame-to-frame standard deviation weighted by the modulus of the bispectrum divided by $\sqrt{50}$). The gray rectangles represent the periods when no data were taken.}
\label{CP_Vega}
\end{figure*}

\begin{table}
\caption{Mean closure phase results and statistical errors.}
\label{CPresults}
\centering
\begin{tabular}{c c c c}
 Baseline Triangle & CP Vega (\degr) & CP Deneb (\degr) & Calibrated CP (\degr)\\
 \hline
1: 1-2-6 & $-10.4\pm0.4$ & $-9.4\pm0.4$ & $-1.0\pm0.6$ \\
2: 1-2-13 & $3.5\pm0.3$ & $1.5\pm0.4$ & $2.0\pm0.5$ \\
3: 1-6-13 & $25.5\pm0.4$ & $23.6\pm1.1$ & $ 1.9\pm1.2$ \\
4: 2-6-13 & $12.2\pm0.4$ & $10.9\pm0.6$ & $1.3\pm0.7$ \\
 \hline
\end{tabular}
\end{table}

\begin{table}
\caption{Comparison of statistical errors. The mean error bars correspond to the average of all individual error bars (shown for each measurement point in Fig. \ref{CP_Vega}). The mean dispersions are the standard deviations of all measurement points.}
\label{CPdispersion}
\centering
\begin{tabular}{c c c c c}
 & \multicolumn{2}{c}{Vega}&\multicolumn{2}{c}{Deneb}\\
 \cline{2-5} 
&&Mean&&Mean\\
Baseline&Mean &dispersion&Mean&dispersion\\
triangle&error bar (\degr)&of data (\degr)&error bar (\degr)&of data (\degr)\\
\hline
1: 1-2-6 & 3.5 & 3.5 & 3.0 & 3.0 \\
2: 1-2-13 & 3.3 & 3.9 & 5.4 & 3.5 \\
3: 1-6-13 & 4.0 & 3.9 & 7.3 & 5.0 \\
4: 2-6-13 & 4.0 & 3.8 & 4.8 & 3.9 \\
\hline
\end{tabular}
\end{table}

\indent The first closure phases are calculated on \object{Vega} using Deneb as a calibrator (Table~\ref{CPresults}). Both are supposed to be unresolved targets with achromatic characteristics. The closure phases are defined by the fibers that form the corresponding triangle. As mentioned above, only four fibers have been used to compute these four closure phases: 1, 2, 6 and 13 are their positions in the v-groove (see Fig. \ref{Pupils}). The longest baseline on the primary mirror is 2.3\,m and corresponds to the pair 6--13.\\
\indent The raw closure phases are presented in Fig.~\ref{CP_Vega}. Each value is computed from the mean of 50 images (10\,s of observation, the exposure time for each image being 200\,ms). Bispectra are estimated for each of the 133 spectral channels and averaged to yield wideband bispectrum values ($\Delta \lambda = 160\,$nm centered on $\lambda_0 = 680\,$nm). Raw statistical error bars for each set of 50 images are first derived from the standard deviation over the 50 measurements weighted by the modulus of the bispectrum. Second, a constant value (the average) is fitted to the sequence of raw measurements and a reduced $\chi^2$ is calculated. If statistical error bars underestimate the fluctuations of the data then the $\chi^2$ is larger than 1. In this case, individual variances are rescaled through a multiplication by the $\chi^2$ value (to get $\chi^2$=1) to derive a more realistic error bar for the closure phase estimates of Table~\ref{CPresults}.

\begin{figure}
\centering
\includegraphics[width=7cm]{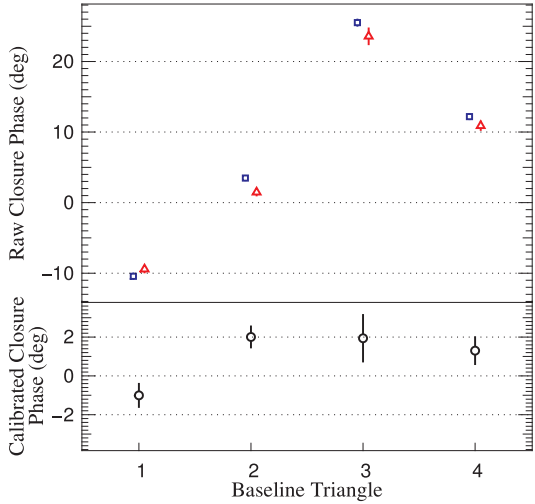}
\caption{\emph{Top:} Closure phase results for \object{Vega} (blue squares) and Deneb (red triangles), which are both supposed to be unresolved targets. The error bars are comparable to or smaller than the symbol size. The exact values are given in Table \ref{CPresults}. \emph{Bottom:} Calibrated closure phase of \object{Vega}. The error bars are the quadratic sum of errors from \object{Vega} and Deneb.}
\label{CPgraph}
\end{figure}

The modulus of the bispectrum, $\vert{\mu_{ijk}}\vert$, is an effective indicator for the detection of the fringes. A signal-to-dark ratio (SDR) has been defined by:
\begin{equation}
SDR_{ijk}=\frac{\vert\mu_{ijk}\vert}{\vert\overline{\mu_{ijk\,dark}}\vert}10^{3\cdot\frac{\,R_\mathrm{mag}}{2.5}},
\end{equation}
with $\vert\overline{\mu_{ijk\,dark}}\vert$ the mean modulus of the bispectrum obtained by applying the pixel to visibility matrix to sky background images, and $R_\mathrm{mag}$ the magnitude of the target. The multiplying factor in the second part of the equation allows for comparable SDR levels for targets with different magnitudes. This quantity does not exactly define a signal-to-noise ratio but constitutes a way to discriminate the data. The data with $SDR\leq 210$ have been rejected; this threshold has been chosen in order to keep around 30\% of the data in the worst case (i.e.\ for triangle 1-6-13 on Deneb). The holes in the sequence of data are due to this selection, especially for Deneb. Moreover, there are gaps with no measurement points that are represented by gray rectangles: they correspond to the optimization of the flux injection into the fibers, requiring about 30 seconds per segment of the mirror (see Sect.~\ref{sec:opti}).\\
\indent Datacubes were taken for different positions of the Dove prism (see Table \ref{Obs}). Rescaled error bars in Fig. \ref{CP_Vega} are comparable to the standard deviation of all measurement points, as shown in Table \ref{CPdispersion}. This means that there is no significant drift or systematic error on timescales of 1--2 hours. In other words, the closure phase measurement is unaffected by mechanical flexures or Dove prism rotation.\\
\indent We have decided to average results for different positions to improve the SDR. The SDR for individual positions was not high enough to detect noticeable effects and we have assumed that 1) instrumental biases would be identical and 2) that any astrophysical signal would be small. The raw closure phases reach high values (up to $26\degr$) and are of instrumental origin. The bispectrum biases due to photon and detector noise could not be properly subtracted because of the lack of calibration data secured during this first run. This is why we have selected only high SDR data to reduce the impact of noise biases. In any case, such biases cannot be the cause of non-zero closure phases.\\
\indent Final calibrated closure phases are obtained by subtracting the Deneb from the \object{Vega} closure phases. The results are presented in Fig. \ref{CPgraph} and listed in Table \ref{CPresults}. The statistical error is on the order of $1\degr$~and at best $0.5\degr$. However, non-zero closure phases are measured at the 2 to 3$-\sigma$ level. It is hard to say at this stage if these are uncalibrated systematic errors or of some astrophysical origin. \object{Vega} is known to be surrounded by a debris disk, which may contribute to visibilities in scattered light. Systematic errors on closure phases are also encountered in aperture masking \citep{Lacour2011} at the 0.3\degr~level and still need to be understood. In our case, they could be due to polarization errors (misalignment of fiber neutral axes, for example) and need to be investigated.

\subsection{Closure Amplitude results}

\begin{figure*}
\centering
\includegraphics[width=17cm]{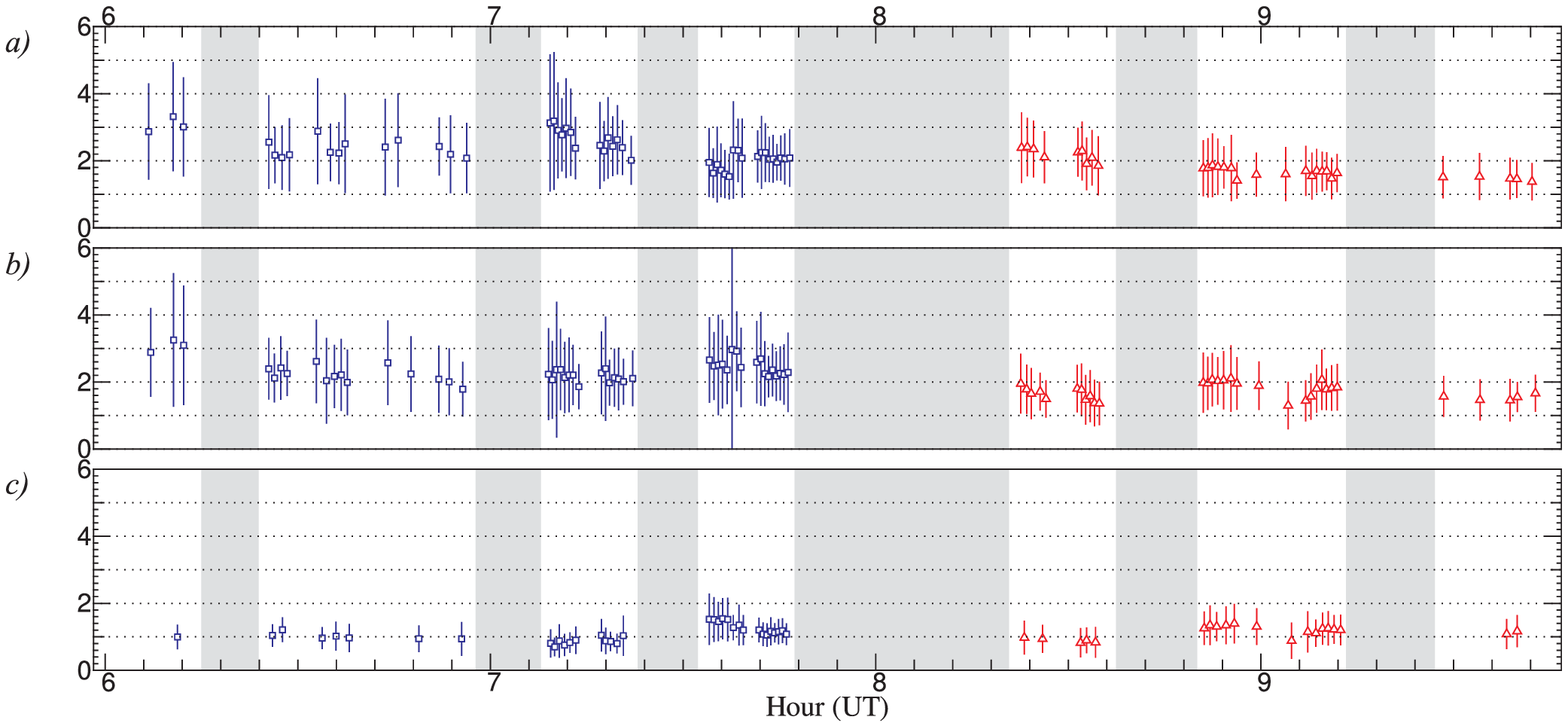}
\caption{Estimates of the closure amplitudes of \object{Vega} (blue squares) and Deneb (red triangles) with rescaled error bars, corresponding to the contributions of four sub-pupils numbered 1, 2, 6 and 13: \emph{a)}: CA$_1$ ; \emph{b)}: CA$_2$ ; \emph{c)}: CA$_3$. Each estimate corresponds to 10\,s of observation (average of 50 images of 200\,ms integrations). The error bars represent the standard deviation of the mean over the 50 frames (i.e. the frame-to-frame standard deviation divided by $\sqrt{50}$). The gray rectangles represent the periods when no data were taken.}
\label{CA}
\end{figure*}

\begin{figure}
\centering
\includegraphics[width=7cm]{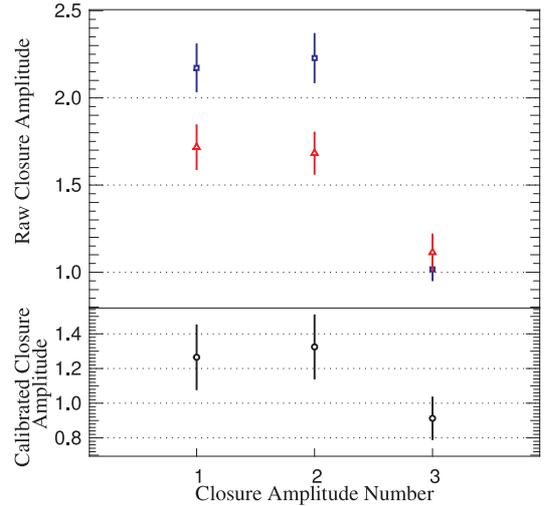}
\caption{\emph{Top:} Closure amplitude results for \object{Vega} (blue squares) and Deneb (red triangles). \emph{Bottom:}Calibrated closure amplitude of \object{Vega}. The error bars are the quadratic sum of errors from \object{Vega} and Deneb.}
\label{CAgraph}
\end{figure}

\indent As explained in Sect.~4.4, visibility amplitudes could not be accurately measured since beam photometries cannot be retrieved without redundancy in the entrance pupil. An order of magnitude can however be estimated assuming typical flux values for the individual beams. The mean total flux per spectral channel is approximately 2$\times$10$^5$ ADU, leading to a typical flux value of 5$\times$10$^4$ ADU per fiber. With a mean coherent flux ranging between 1$\times$10$^4$ and 4$\times$10$^4$ ADU, we estimate the raw contrasts for the 6 baselines to roughly range between 10\,\% and 40\,\%.\\
\indent Closure amplitudes (CA) have been measured and are presented in Fig. \ref{CA}. These estimates have been obtained by following the same procedure as for the closure phase measurements: computing estimates by selecting the data according to a given threshold, estimating the mean over all points, and correcting the final error by the $\chi^2$ calculation. The closure amplitudes are numbered from 1 to 3:
\begin{equation}
CA_1=\left| \frac{\mu_{2-1} \mu_{13-6}}{\mu_{6-1}\mu_{13-2}} \right|, \\
CA_2=\left| \frac{\mu_{2-1} \mu_{13-6}}{\mu_{13-1}\mu_{6-2}} \right|, \\
CA_3=\left| \frac{\mu_{6-1} \mu_{13-2}}{\mu_{13-1}\mu_{6-2}} \right|.
\label{eq:CA} 
\end{equation}
Fig. \ref{CA} shows that the closure amplitudes are stable over about half an hour. The observation procedure could therefore be optimized by alternating the target and the calibrator observations more frequently.\\
\indent The calibrated closure amplitudes of Vega are shown in Fig. \ref{CAgraph}. They are close to the expected value of unity for these unresolved targets. CA$_1$ and CA$_2$ are measured with precisions of about 25\,\%, which is not sufficient for high dynamic range detection. According to Eq. \ref{eq:CA}, these closure amplitudes have 2 baselines in common: (2-1) and (13-6). The biases observed in the resulting closure amplitudes could possibly come from the contrast measurements of these bases, but as for closure phases, we must further investigate the sources of these biases.\\
\indent On the other hand, CA$_3$ is stable: dispersion of the data points in Fig. \ref{CA} is 0.2 whereas the overall mean error bar is 0.4. Moreover, the estimates of the mean are 1.0 for Vega and 1.1 for Deneb, with accuracies of 10\,\%. In other words, the third closure amplitude is fully consistent with a point source nature for both objects and a proper data analysis.

\subsection{Dynamic range}

Given the results presented in the previous Section, only the closure phases will be used to evaluate the dynamic range detection limits.\\
\indent \citet{Lacour2011} gives a 1$\sigma$ dynamic range estimation for sparse aperture masking using 7 sub-pupils, as a function of the error on closure phase measurements based on a Monte Carlo simulation:
\begin{equation}
1\sigma\,\mathrm{dynamic~range}=\frac{400}{\sigma(CP)_{\mathrm{degr}}}.
\end{equation}
This equation can be extrapolated to estimate the 4$\sigma$ dynamic range for any number of baselines $n_\mathrm{B}$ (based on the theoretical derivation by \citet{Baldwin2002}, who have established that the dynamic range increases with the square root of the number of baselines):
\begin{equation}\label{eq:dyn}
4\sigma\,\mathrm{dynamic~range} \sim 22 \frac{\sqrt{n_\mathrm{B}}}{\sigma(CP)_{\mathrm{degr}}}.
\end{equation}

\indent The calibrated closure phase estimations presented in this first paper reach a final accuracy of 0.75\,\degr\,on average, resulting from 17\,min total integration time on Vega and 12\,min on Deneb. This leads to a 4$\sigma$ dynamic range of 180 in the reconstructed image, if all 9 fibers were equally transmissive.

\subsection{Sensitivity}

\begin{figure}
\centering
\includegraphics[width=7cm]{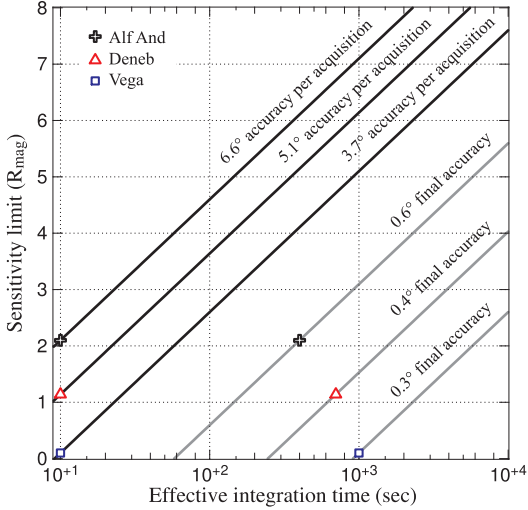}
\caption{Sensitivity limits as a function of the integration time for given closure phase accuracies per acquisition. These curves have been extrapolated from the accuracies measured on Vega (square at R$_{mag}$=0.1), Deneb (triangle at R$_{mag}$=1.14) and Alf And (cross at R$_{mag}$=2.1). Two curves are drawn per target: the black one is extrapolated from the mean accuracy per acquisition and the gray one from the final best statistical error. The accuracies are given for uncalibrated measurements.}
\label{Sensitivity_limit}
\end{figure}

During the first run, the target list was limited to bright objects. Nevertheless, the sensitivity of the instrument can be estimated by extrapolating the results obtained on the faintest target, which is Alf And (B9II star, V$_\mathrm{mag}$=2.06, R$_\mathrm{mag}$=2.1), observed under the same conditions (200\,ms integration time). The same procedure has been applied to the 4 sets of 10 datacubes acquired on this target to obtain closure phase measurements. The statistical error bar of one acquisition (resulting from 50 200ms-images) is 6.6\,\degr\, on average, and 4.4\,\degr\,for the triangle with the highest signal-to-noise ratio. When combining all data, the final statistical error is 0.6\,\degr\, for 7\,min effective integration time.\\
\indent Sensitivity limits can be extrapolated from the results obtained on the different targets and can be derived for a given closure phase accuracy, measured on targets with R$_{mag\,0}$ and integration time $\tau_0$:
\begin{equation}
R_{mag\,lim}=R_{mag\,0}+2.5 \log \left(\frac{\tau}{\tau_0} \right).
\end{equation}
These limits are derived under the assumption of the photon noise limited regime. A signal-to-noise ratio of at least 1 is needed per exposure and per spectral channel in order for the overall integrated signal-to-noise ratio to be reasonable. In other words, at least one photon is needed per fiber per frame per coherence time. As FIRST is working behind an AO system, the coherence time is theoretically infinite and integration times could be set to seconds or more for the faintest targets.\\
\indent Sensitivity limits are plotted in Fig. \ref{Sensitivity_limit} as a function of effective integration time for one acquisition (50$\times\tau_0$) and a given accuracy. These accuracies are for uncalibrated measurements, meaning that the final accuracies should be multiplied by a factor $\sqrt{2}$ in the best case (reference target of the same magnitude as the science target), and effective integration times should be doubled to take the reference target observations into account. The current state of the prototype could measure closure phases of targets at R$_{mag}\leq4$ with a 3.7\,\degr\,accuracy per acquisition (uncalibrated) in about 6\,minutes, and with a 0.6\,\degr\,accuracy in 40\,min. The achievable magnitude is higher for relaxed constraints on the accuracy.

\section{Discussion}

\label{sec:discussion}

The first on-sky results with FIRST demonstrate that it is possible to operate this type of instrument with multiple single-mode fibers. Several key ingredients have been used: the segmented mirror that precisely steers beams into fibers, and the microlens arrays, and fiber bundle which compactly focus light into the fibers. However, the throughput is low and the accuracy of the closure phase measurements are currently limited by systematic errors. Future developments of the FIRST instrument will be dedicated to improving these aspects.
\subsection{Sensitivity and robustness: Long-term improvements}
Given the throughput budget presented in Table~\ref{Transmissions}, some improvement can be made by decreasing the number of optical surfaces to reach a higher theoretical efficiency (currently a few percent). \\
\indent This could be achieved in several ways. First of all the optical design could be simplified insofar as it could be better integrated into the AO system. Indeed the prototype set-up is complex because the AO and FIRST are separate systems. For instance, the mirrors required to control the pupil and image planes in FIRST (M1 and M2) are redundant with their counterparts in the AO system. This is the same for the optical train required to extract the beam from the AO system (one beam splitter cube and two mirrors, which are not drawn in Fig. \ref{Setup}, are currently needed). The set-up could also be simplified by placing the segmented mirror at quasi normal incidence. This would cause negligible flux losses compared to the gain in throughput: the polarizing beam splitter cube and the quarter wave plate could be removed (removing 10 reflections/transmissions). In this case, the control of the image in the field of view would not be done continuously but alternately, by putting mirror M1 or M2 on a translation stage to let the beam reach the alignment CCD.\\
\indent Second, part of the output optics of FIRST could be replaced by a 3D integrated photonics pupil remapper \citep[as for the Dragonfly instrument,][]{Tuthill2010}, avoiding the fiber-interfacing between the fiber bundle and the v-groove, and providing a more robust and thermally stable instrument. Integrated optics could also replace some of the recombination optics, such as the anamorphic system (see \citet{Benisty2009} for 1D integrated optics and \citet{Minardi2010} for 2D integrated optics). The development of these technologies is in progress. Currently only prototypes working in the near-infrared wavelength range have been tested on-sky, while integrated optics at visible wavelengths remain to be developed.

\subsection{Next prototype: FIRST-18}
In order to reach better performance and allow for observations of astrophysical interest in the shorter term, three axes of improvement must be implemented on the current prototype. A major identified limitation to the measurement accuracy is mechanical stability. The prototype was first made to be a laboratory instrument, yet was mounted directly on the Cassegrain focus of the Shane telescope. Mechanical instabilities such as flexure have turned out to be a critical issue during the first run. The bench will undergo substantial improvements to provide better injection into fibers and therefore increase the sensitivity. Second, the next version of the prototype, named FIRST-18, will consist of 18 fibers, thus increasing the number of available baselines. The pupil mapping will be updated to avoid the central obstruction of the telescope. Finally, better protection from dust and a better coating of the optics will ensure a higher throughput to reach the 1\,\% theoretical efficiency.\\
\indent As stated by Eq. \ref{eq:dyn}, the dynamic range will be increased due to the larger number of fibers and improved sensitivity and mechanical stability, enabling more accurate closure phase measurements. The current accuracy on closure phases may be primarily limited by systematics which will be reduced by data processing and by better calibration procedures.

\section{Conclusion}

We have obtained the first on-sky results with FIRST, a fibered aperture masking instrument that aims at providing a unique combination of high contrast and high angular resolution. Although the described performances are not very high, better precisions and sensitivity are expected in a near future.\\
\indent Apart from the systematic errors that may remain, we have measured calibrated closure phases on Vega with an accuracy of 0.5\,\degr\ in the best case by co-adding $\sim$\,100 acquisitions (with accuracies around 3-4\,\degr\,per acquisition of 10\,s), totaling 17\,min exposure time. This leads to a 4$\sigma$ dynamic range of around 260 in a reconstructed image when considering a 9-fiber prototype. This dynamic range would increase with the square root of the number of acquisitions \citep{Perrin2006}, if not limited by systematic errors. By extrapolation, the 4$\sigma$ dynamic range could therefore be on the order of $10^3$ when co-adding 14 17min-acquisitions, equaling about 4\,hours total effective integration time for both target and calibrator. However if we achieve a closure phase accuracy of 0.4\,\degr\,in 17\,min, the same dynamic range could be achieved in 2.6\,hours integration time. Moreover, when the 30-fiber instrument will be completed, the number of baselines will be 210 (if we assume 2 recombination paths of 15 fibers), and a dynamic range of 10$^4$ could be reached in less than 3 hours if a bias-corrected closure phase accuracy of 0.1\,\degr\, is achieved in 17\,min.\\
\indent Closure amplitudes have not been used to evaluate the dynamic range detection limits because of the limited precision on these estimates (10\,\% in the best case). With a fully working instrument, data processing would allow the computation of the visibilities, leading to better performance in high dynamic range detection.\\
\indent The preliminary data presented in this paper, although they do not show very high performances, constitute an important demonstration of the fibered pupil remapping concept applied to a monolithic telescope. FIRST in the visible range is not competitive resolution-wise with long baseline interferometers. Long-baseline instruments provide much higher angular resolution in the visible, but with more limited field of view, unless spectral resolutions of typically a few thousands are used. FIRST on 3 to 10--m class telescopes is a good complement, as it gives access to a range of large objects and to spatial frequencies otherwise unaccessible with either IR-optimized adaptive optics systems or long baseline interferometers. Moreover, a fibered aperture masking instrument is also complementary in terms of dynamic range since it can provide high dynamic range at the diffraction limit, while AO systems can reach high dynamic range at a few $\lambda/D$. Possible science applications are stellar surfaces for giants and supergiants, circumstellar environments, debris disks, etc ... Although FIRST could eventually be used to observe the closest and brightest planets, a near-infrared version would be better suited to exoplanetary system studies.\\
\indent Our future efforts in the development of FIRST will be concentrated on the improvement of the sensitivity along with the accuracy of the closure phases and visibilities. Dynamic ranges up to 10$^3$-10$^4$ and a sensitivity limit of R$_{mag}\geq4$ seem within range in the not too distant future.

\begin{acknowledgements}
The authors would like to thank Dr. Bolte, Director of the University of California Observatories for his commitment and financial support, as well as the Lick Observatory staff for its precious help during the runs. Special thanks go to Dr. Helmbrecht, President and Founder of Iris AO who built the segmented mirror for his strong support. F.M. work was supported by the National Science Foundation under award number AAG-0807468 and through a private donation through the ``Adopt a Scientist program'' at the SETI Institute by Jeff Breidenbach and Jeff Marshall. We acknowledge financial support from ``Programme National de Physique Stellaire'' (PNPS) of CNRS/INSU, France.
\end{acknowledgements}

\bibliographystyle{aa}
\bibliography{FIRST_biblio}

\end{document}